\documentclass[%
 reprint,
 amsmath,amssymb,
 aps,
]{revtex4-1}

\usepackage{graphicx}
\usepackage{dcolumn}
\usepackage{bm}

\usepackage{booktabs}

\begin{document}

\preprint{APS/123-QED}

\title{A Method for Measuring the $6S_{1/2} \leftrightarrow 5D_{3/2}$ Magnetic Dipole Transition Moment in {Ba}$^{+}$ }

\author{Spencer R. Williams }
 \email{srw8@u.washington.edu}
\affiliation{%
 Department of Physics, University of Washington, Seattle, Washington, 98195, USA\\
 }%
\author{Anupriya Jayakumar}%
\affiliation{%
 Department of Physics, University of Washington, Seattle, Washington, 98195, USA\\
}%

\author{Matthew R. Hoffman}%
\affiliation{%
 Department of Physics, University of Washington, Seattle, Washington, 98195, USA\\
}%

\author{Boris B. Blinov}%
\affiliation{%
 Department of Physics, University of Washington, Seattle, Washington, 98195, USA\\
}%

\author{E.N. Fortson}%
\affiliation{%
 Department of Physics, University of Washington, Seattle, Washington, 98195, USA\\
}%

\date{\today}

\begin{abstract}
	    \noindent We propose a method for measuring the magnetic dipole (M1) transition moment of the $6S_{1/2} \big( \mathrm{m}=-1/2\big)\leftrightarrow 5D_{3/2}\big( \mathrm{m}=-1/2\big)$ transition in single trapped Ba$^{+}$ by exploiting different symmetries in the electric quadrupole (E2) and M1 couplings between the states.  The technique is adapted from a previously proposed method for measuring atomic parity nonconservation in a single trapped ion [Norval Fortson, Phys. Rev. Lett. \textbf{70}, 17 (1993)].  Knowledge of M1 is crucial for any parity nonconservation measurement in Ba$^{+}$, as laser coupling through M1 can mimic the parity-violating signal.  The magnetic moment for the transition has been calculated by atomic theory and found to be dominated by electron-electron correlation effects [B.K. Sahoo et. al., Phys. Rev. A \textbf{74}, 6 (2006)].  To date the value has not been verified experimentally.  This proposed measurement is therefore an essential step toward a parity nonconservation experiment in the ion that will also test current many-body theory.  The technique can be adapted for similar parity nonconservation experiments using other atomic ions, where the magnetic dipole moment could present similar complications.
	     
\begin{description}
\item[PACS numbers] 31.30.jc, 31.30.jg, 32.70.Cs, 32.70.Fw	
\end{description}
\end{abstract}

\pacs{Valid PACS appear here}
\maketitle

\section{Introduction}

In atoms the parity nonconserving (PNC) exchange of the Z$_{0}$ neutral gauge boson between the nucleons and electrons manifests itself as a non-vanishing electric dipole transition moment ($\mathrm{E1_{PNC}}$) between states of the same parity \cite{Bouchiat74, Bouchiat75}.  Precision measurements of $\mathrm{E1_{PNC}}$, together with high accuracy atomic theory calculations, offer a unique opportunity to study electro-weak physics in a low energy experiment with the possibility to probe for physics beyond the standard model \cite{Marciano90, Fortson90, Fortson92, Haxton08, Blundell92}.  To date the most accurate measurement of $\mathrm{E1_{PNC}}$ was performed on a beam of atomic cesium which achieved an experimental uncertainty of 0.39\% \cite{Wood97}.  Combined with the most recent atomic theory \cite{Derevianko00,Porsev09,Dzuba12}, the result agrees with the standard model to within 1.5\,$\mathrm{\sigma}$.  Also, at this level of uncertainty the authors were the first to observe a nuclear anapole moment.

\begin{figure}[t]
\includegraphics[scale=0.52]{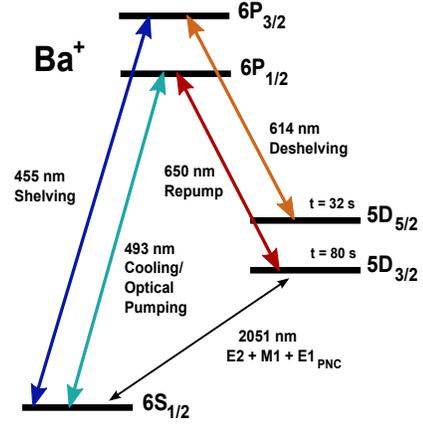}
\caption{(Color online) A partial energy diagram for $\mathrm{Ba^{+}}$ showing its lowest laying states and the transitions that are relevant to the M1 measurement.  The 2051 nm transitions, on which we will focus our attention, have contributions from both electric quadrupole and magnetic dipole transition moments.  Parity nonconservation also induces a very small electric dipole transition moment $\mathrm{E1_{PNC}}$ to the $5D_{3/2}$ states.  The other transitions are strongly connected by electric dipole moments; higher order moments along these transitions are inconsequential to the measurement proposed here.}
\label{fig:barium}
\end{figure}

\indent  One promising proposal for future PNC measurements, with comparable or better experimental accuracy, 
seeks $\mathrm{E1_{PNC}}$ through precision spectroscopy of heavy single trapped ions \cite{Norval93}. This approach has the potential to reach experimental uncertainties below $0.1\%$ and is currently being pursued in Ba$^{+}$ \cite{Sherman05} , Yb$^{+}$ \cite{Rahaman13}, and Ra$^{+}$ \cite{Versolato11}.  Such experiments will measure the modulation of the Rabi frequency $\mathrm{\Omega}$ for a $\big(n\big)S_{1/2} \leftrightarrow \big(n-1\big)D_{3/2}$ transition (see Fig \ref{fig:barium}.) due to the non-zero interference between the electric quadrupole (E2) transition moment and $\mathrm{E1_{PNC}}$,

\begin{equation}\label{eq:PNCRabi}
\mathrm{\Omega}^{2} = \big| \mathrm{\Omega_{E2}} + \mathrm{\Omega_{\mathrm{PNC}}} \big|^{2} \approx \mathrm{\Omega^{2}_{E2}} \pm 
2 \mathrm{Re}\big( \mathrm{\Omega_{E2}}\mathrm{\Omega^{*}_{PNC}} \big),
\end{equation} 
\noindent where $\mathrm{\Omega_{E2}}$ and $\mathrm{\Omega_{PNC}}$ are the E2 and E1$_{PNC}$ contributions to the total Rabi frequency ($\mathrm{\Omega}$). The parity violating $\mathrm{E1_{PNC}}$ can be extracted from the interference term in Eq. (\ref{eq:PNCRabi}) \,.    
The experiment is complicated by the existence of a non-vanishing magnetic dipole (M1) moment between the same states. In particular, for Ba$^{+}$  the apposite reduced transition moments are,

\begin{subequations}\label{eq:M1E2}
\begin{equation}
\mathrm{M1} = \big\langle 6S_{1/2} \big|\big| \widehat{\mathrm{M1}} \big|\big| 5D_{3/2} \big\rangle
\end{equation}
\begin{equation}
\mathrm{E2} = \big\langle 6S_{1/2} \big|\big| \widehat{\mathrm{E2}} \big|\big| 5D_{3/2} \big\rangle
\end{equation}
\end{subequations}

\noindent Where $\widehat{\mathrm{M1}}$ and $\widehat{\mathrm{E2}}$ are the magnetic dipole and electric quadrupole operators, respectively.  Coupling through the M1 moment can mimic the interference term in Eq.(\ref{eq:PNCRabi}), making it a potentially serious systematic problem for any trapped ion PNC experiment \cite{Mandal10}.  Calculations of M1 and E2 were recently reported \cite{Sahoo06} and are given in Table \ref{tab:relativesizes} along with an estimation of E1$\mathrm{_{PNC}}$ for scale.  This work predicts an M1 dominated by electron-electron correlation effects, but the value has yet to be corroborated.  A measurement of M1 would be an important test of many-body theory and is an essential step toward a PNC experiment in Ba$^{+}$.  In the present work we describe a method for measuring M1 that exploits the presence of the significantly larger E2 moment. \\

\begin {table}[ht]
\begin{center}
\begin{tabular}{ >{\centering\arraybackslash}m{1.25in}  >{\centering\arraybackslash}m{.85in} >{\centering\arraybackslash}m{.75in} >{\centering\arraybackslash}m{.75in} >{\centering\arraybackslash}m{.75in} >{\centering\arraybackslash}m{.75in}}
\toprule[1.5pt]
{\bf $\mathrm{E2}$} & {\bf $\mathrm{M1}$} & {\bf $\mathrm{E1_{PNC}}$}  \\ 
\hline 
12.6 ($\frac{a_{0}}{\lambda}$) & 8.0 $\times$ 10$^{-4}$ ($\frac{\alpha}{2}$) & $\sim$ 2$\times$ 10$^{-11}$ \\
$\sim$ 1               & $\sim$ 10$^{-2}$      & $\sim$ 10$^{-7}$     \\
\bottomrule[1.25pt]
\end {tabular}
\caption {Calculated \cite{Sahoo06,Safronova13,Sahoo07,Dzuba06,Roberts13} and relative sizes of the transition moments between $6S_{1/2}$ and $5D_{3/2}$ in Ba$^{+}$.  The calculated values are listed in units of ea$_{0}$, where e is the electron charge and a$_{0}$ is the Bohr radius.  The fine structure constant $\mathrm{\alpha}$ and the transition wavelength $\mathrm{\lambda}$ enter so that a \emph{bona fide} comparison can be made with these units.  Although the calculation for M1 shows enhancement from electron-electron correlation effects it is still small relative to E2.}
  \label{tab:relativesizes}
\end{center}
\end {table}

\subsection{Measurement of Rabi Frequencies in Ba$^{+}$}
\indent The M1 measurement proposed in this work assumes the ability to determine several Rabi frequencies for transitions to particular Zeeman sub-levels of the $5D_{3/2}$ manifold from the $6S_{1/2}$ ground states.  In this section we describe the shelving technique \cite{Nagourney86} which can be used to measure all of the required Rabi frequencies.  Here the technique is presented with the use of lasers at 455 nm and 614 nm to drive the transitions indicated in Fig.\ref{fig:barium}\,.  The method exploits the very long lifetimes of the $5D_{3/2}$ and $5D_{5/2}$ states, which are 80 s \cite{Yu97} and 32 s \cite{Nagourney86}, respectively. 

\begin{figure}[ht]
\includegraphics[scale=1.03]{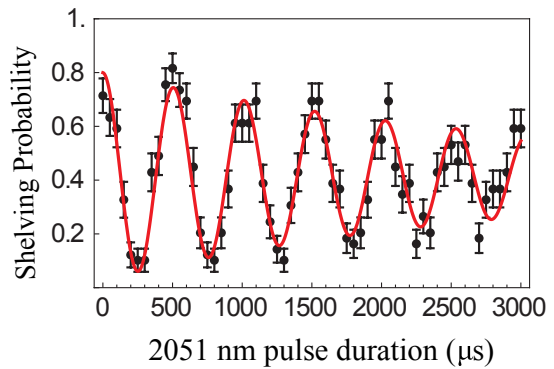}
\caption{(Color online) Rabi oscillations on the $6S_{1/2}\big( \mathrm{m}=-1/2\big)$ to $5D_{3/2}\big( \mathrm{m}=-3/2\big)$ transition reported in \cite{Hoffman12}.  A Rabi frequency of 2 kHz was obtained with a decoherence rate of about 300 Hz.}
\label{fig:RabiFlopCrop}
\end{figure}

\indent Ba$^{+}$ is laser cooled along the $6S_{1/2} \leftrightarrow 6P_{1/2}$ transition with 493 nm light.  A repump laser at 650 nm is required because of the significant branching ratio for spontaneous decay to the $5D_{3/2}$ states from $6P_{1/2}$ \cite{Kurz08}.  Fluorescence at 493 nm is collected so that the ion can be observed while it cycles on the 493 nm and 650 nm transitions.  From $6S_{1/2}$ the ion can be pumped to the $5D_{5/2}$ state by a pulse of 455 nm light.  Since an ion in this state has been removed from the cooling cycle no photons will be emitted with application of the cooling beams and the ion is said to be ``shelved".  Conversely, if the ion is initially driven to the $5D_{3/2}$ state then the 455 nm laser will be unable to shelve the ion and it will fluoresce when addressed by the cooling beams.  This generates a binary signal, in the form of a ``bright" or ``dark" ion, that indicates which state the ion occupies.  The following pulse sequence uses this signal to determine the Rabi frequency for any of the 2051 nm transitions.\\
\indent After the ion is initially cooled, the 493 nm and 650 nm lasers are turned off and the ion state is initialized to $6S_{1/2}\big( \mathrm{m}=-1/2\big)$ by optical pumping with circularly polarized 493 nm light.  One then attempts to drive the ion to a particular $5D_{3/2}$ Zeeman sub-level by delivering a pulse of resonantly tuned 2051 nm light for a time $\tau$, after which the ion will have some probability of being found in that particular $5D_{3/2}$ sub-level.  A 455 nm shelving pulse is then delivered and the cooling lasers are used to interrogate whether the attempt to shelve the ion was successful.  A 614 nm pulse returns the ion to the ground state at the end of the sequence if necessary.  This pulse sequence is repeated until the probability the ion was shelved, $\mathrm{P_{s}(\tau)}$, is determined to the desired uncertainty.  Depending on the size of the Rabi frequency $\mathrm{\Omega}$ compared to the experiment's decoherence rate $\mathrm{\gamma}$ the shelving probability can take two forms, 
 
\begin{subequations}\label{eq:ShelveEff}
\begin{equation}\label{eq:ShelveEffA}
\begin{split}
\mathrm{P_{s}(\tau)}=&\mathrm{\frac{\epsilon}{2}\Big[1 + e^{-\gamma\tau}\big(\,\cos(\Omega^{'}\tau) + \frac{\gamma}{\Omega^{'}}\sin(\Omega^{'}\tau)\,\big) \Big] }\\
&\mathrm{when}\quad\mathrm{\Omega>\gamma}\quad\mathrm{(underdamped)}
\end{split}
\end{equation}
\begin{equation}\label{eq:ShelveEffB}
\begin{split}
\mathrm{P_{s}(\tau)} =  &\mathrm{\frac{\epsilon}{2}\Big[1 + e^{-\gamma\tau}\big(\,\cosh(\gamma^{'}\tau) + \frac{\gamma}{\gamma^{'}}\sinh(\gamma^{'}\tau)\,\big)\Big]}\\
&\mathrm{when}\quad\mathrm{\gamma>\Omega} \quad\mathrm{(overdamped)}
\end{split}
\end{equation}
\end{subequations}

\noindent where $\mathrm{\Omega^{'} = \Omega\sqrt{1-\big(\gamma/ \Omega\big)^{2}}}$ and $\mathrm{\gamma^{'} = \gamma\sqrt{1-\big(\Omega/\gamma\big)^{2}}}$.  For this procedure, the maximum shelving efficiency $\mathrm{\epsilon}$ will be less than unity and is theoretically limited to 0.87 by the branching ratio for spontaneous decay to $5D_{3/2}$ from $6P_{3/2}$ \cite{Kurz08}.  Spectroscopy of the $5D_{3/2}$ states was recently reported with a frequency stabilized 2051 nm laser \citep{Hoffman12} using the shelving technique described here.  An example of the shelving probability plotted against $\mathrm{\tau}$ is shown in Fig. \ref{fig:RabiFlopCrop} for the transition between $6S_{1/2}\big( \mathrm{m}=-1/2\big)$ and $5D_{3/2}\big( \mathrm{m}=-3/2\big)$. In that work a Rabi frequency of 2 kHz was achieved with a decoherence rate around 300 Hz due largely to ambient magnetic field drift.  The technique to measure M1 that follows necessitates driving transitions at relatively low Rabi frequencies compared to what was found in \citep{Hoffman12} which suggests the importance of minimizing sources of decoherence.  

\section{M1 Measurement Procedure}
\indent In principle, if the magnetic dipole transition moment's contribution to the total Rabi frequency is known then that transition moment can be extracted if the driving field's alignment, intensity, and polarization are known at the ion.  In the odd isotope, $^{137}\mathrm{Ba}^{+}$, the E2 amplitude  vanishes for $\mathrm{F=1} \rightarrow \mathrm{F=0}$ transitions, allowing for a pure magnetic dipole transition.  However, a direct measurement is unfavorable because of the the modest size of M1 and complications with working in the odd isotope.  In $^{138}\mathrm{Ba}^{+}$,  coupling via E2 is not suppressed but the Rabi frequency can be modulated with experimental parameters so as to isolate $\mathrm{\Omega_{M1}}$.  In the parameter space of interest, to be defined in the forthcoming discussion, the ratio of the relative contributions to the Rabi frequency from each moment, $\mathrm{\Omega_{M1}}$/ $\mathrm{\Omega_{E2}}$, will be of order 0.1\,.  From the Rabi frequency reported in \cite{Hoffman12}, we estimate that $\mathrm{\Omega_{M1}}$ will be tens of hertz. 

\begin{figure}[ht]
\includegraphics[width=84 mm]{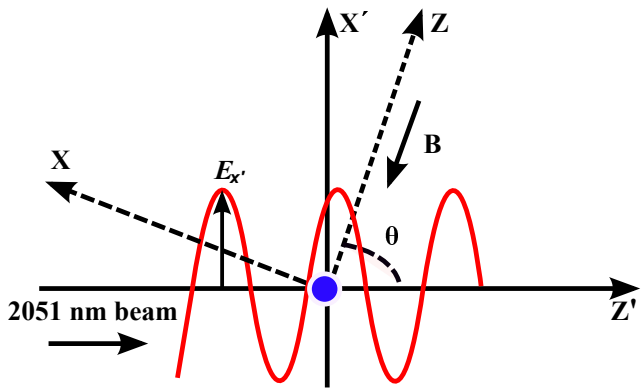}
\caption{(Color online) The primed coordinate system defines the laboratory frame with the Z$^{'}$-axis set along the 2051 nm laser.  The unprimed coordinates define the atom's frame with the Z-axis set along the externally applied magnetic field $\mathrm{\textbf{B}}$.  The Y and Y$^{'}$ axes are parallel and point out of the page.}
\label{fig:coor}
\end{figure}
 
\indent The controlled modulation of the Rabi frequency can be most simply performed on the $6S_{1/2}\big( \mathrm{m}=-1/2\big)\leftrightarrow 5D_{3/2}\big( \mathrm{m}=-1/2\big)$ transition.  A quantization axis to lift the degeneracy between the Zeeman sub-levels is established with a static magnetic field applied by a pair of static current carrying coils in a Helmholtz-like configuration.  We define $\theta$ to be the angle between the applied magnetic field and the 2051 nm beam, as depicted in Fig. \ref{fig:coor}.  The Rabi frequency for the $\mathrm{\Delta  m=0}$ transition can then be written in terms of its E2 and M1 contributions as,

\begin{subequations}\label{eq:Rabiform}
\begin{equation}\label{eq:RabiformA}
\mathrm{\Omega} = \big|\mathrm{\Omega_{E2}} + \mathrm{\Omega_{M1}} \big|
\end{equation}
\begin{equation}\label{eq:RabiformB}
\mathrm{\Omega_{E2}} = \frac{i k}{4 \hbar}\sqrt{\frac{1}{10}} \mathrm{E2}\,\sin(2\theta)\,\mathrm{E}_{x^{'}}
\end{equation}
\begin{equation}\label{eq:RabiformC}
\mathrm{\Omega_{M1}} =  -\frac{1}{\hbar}\sqrt{\frac{1}{6}}\mathrm{M1}\,\sin(\theta)\,\mathrm{B}_{x^{'}}
\end{equation}
\end{subequations}

\begin{figure}[t]
\includegraphics[scale=0.33]{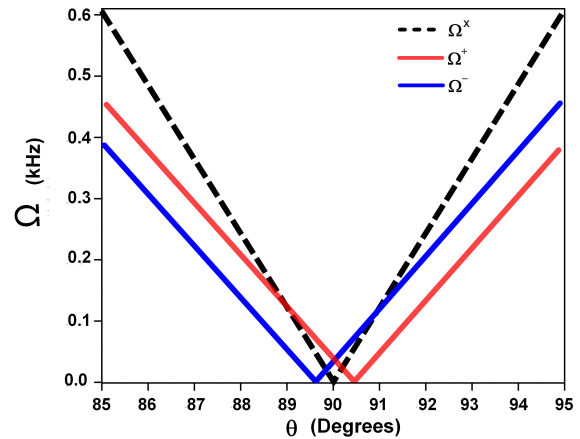}
\caption{(Color online) The solid blue and red curves are $\mathrm{\Omega^{+}}({\theta})$ and $\mathrm{\Omega^{-}}({\theta})$ respectively, where the $\pm$ reflects the handedness of the 2051 nm laser polarization.  The dashed curve shows $\mathrm{\Omega^{x}}(\theta)$, which is the same transition driven by horizontally polarized light, that could be useful for calibrating $\mathrm{\theta}$.  All numerical values are estimated using theoretical values \cite{Sahoo06} for the transition moments and an electric field intensity estimated from previous measurements \citep{Hoffman12}}
\label{fig:rabitheta}
\end{figure}

\noindent Where $\mathrm{E}_{x^{'}}$ and $\mathrm{B}_{x^{'}}$ refer to the components of the 2051 nm laser beam fields.  To have both $\mathrm{\Omega_{E2}}$ and $\mathrm{\Omega_{M1}}$ be non-zero and add in-phase the transition should be driven with circularly polarized light. A plot of the expected value of $\mathrm{\Omega}$ driven by either sense of circular polarization and linear light is shown in Fig. \ref{fig:rabitheta}. 

\indent It is evident in Eq. (\ref{eq:Rabiform}) that $\mathrm{\Omega_{M1}}$ and $\mathrm{\Omega_{E2}}$ possess even and odd symmetry, respectively, about $\theta = 90^{\circ}$.  The $\mathrm{\Omega_{E2}}$ contribution to $\mathrm{\Omega}$ can be canceled by symmetrically shifting $\mathrm{\theta}$ about 90$^{\circ}$ by a small angle $\mathrm{\delta}$ as

\begin{equation}\label{eq:M1inter}
\mathrm{\Delta \Omega} =\big| \mathrm{\Omega}(90^{\circ}+\delta) - \mathrm{\Omega}(90^{\circ}-\delta) \big| = 2\mathrm{\Omega_{M1}}
\end{equation}

\noindent with $\mathrm{\delta}$ chosen to be a few degrees.  The value of $\mathrm{\Omega\big( \theta \big)}$ can be extracted from a decay curve as illustrated in Fig. \ref{fig:RabiFlopCrop}.  An accurate determination of $\mathrm{\Omega_{M1}}$ from $\mathrm{\Delta \Omega}$ requires precise tuning of $\theta$.  We estimate that to measure $\mathrm{\Omega_{M1}}$ to five percent accuracy each orientation must be known to within 0.02$^{\circ}$.  The offset angle $\mathrm{\delta}$ can be tuned by rotating the magnetic field coils about the trap center or by adjusting a second set of orthogonally placed coils, but imperfections in the magnetic field make it difficult to know precisely how much they need to be adjusted \emph{a priori}.  It is therefore crucial that $\mathrm{\theta}$ be calibrated for each measurement.

\indent None of the 5$D_{3/2}$ transitions are useful for such delicate angular calibration if driven with circularly polarized light.  However, a suitable configuration could be to drive the $\mathrm{\Delta}$m = 0 transition with horizontally polarized light (electric field parallel to the X$^{'}$-axis) for which the Rabi frequency, $\mathrm{\Omega^{X}(\theta)}$, is sharply peaked and symmetric about 90$^{\circ}$, as shown in Fig. \ref{fig:rabitheta}.  For the two orientations of $\mathrm{\theta}$, an agreement between either $\mathrm{\Omega^{X}(90^{\circ}\pm \delta)}$ to 1\% or better is sufficient to calibrate $\mathrm{\theta}$.  This may be challenging if operating in the overdamped regime (Eq. \ref{eq:ShelveEffB}).  Because the $\mathrm{\Delta m=0}$ Rabi frequency is expected to be small, it will be useful to reduce $\mathrm{\gamma}$ as much as is practical.  Since the 300 MHz decoherence rate in \cite{Hoffman12} was mostly due to magnetic field noise in an unshielded setup, we expect significant improvement to $\mathrm{\gamma}$ with the use of magnetic shielding.

\indent An alternative method to modulate $\mathrm{\Omega}$, using only one position of $\theta$, is to drive the transition with both senses of circular polarization.  In this approach $\mathrm{\Omega_{M1}}$ retains the proper relative phase and changes sign exactly as in Eq. (\ref{eq:M1inter}) but with respect to the handedness of the 2051 nm beam, indicated by a + or - superscript:
	
\begin{equation}\label{eq:M1pol}
\mathrm{\Delta \Omega} = \big| \mathrm{\Omega^{+}}(90^{\circ}\pm\delta) - \mathrm{\Omega^{-}}(90^{\circ}\pm\delta)\big| = 2\mathrm{\Omega_{M1}}
\end{equation}	

\noindent In this approach the error in the relative positioning of $\theta$ between either measurement can be limited to the stability of the current source driving the magnetic field coils.  Here then $\theta$ need only be known to $\mathrm{\sim 1^{\circ}}$, primarily to ensure that $\mathrm{\Omega_{M1}}$ is approximately maximized.  Care must be taken, particularly with this approach, to have clean circular polarization in the 2051 nm beam.  Systematic distortions to the polarization can, in principle, be compensated, however it will be desirable to limit distortions to a few percent.

\indent  To extract the M1 moment from Eq. \ref{eq:RabiformC}, the 2051 nm light field at the ion must be measured.  This can be done with the Rabi frequency of the $6S_{1/2}\big( \mathrm{m}=\mp 1/2\big)\leftrightarrow 5D_{3/2}\big( \mathrm{m}=\pm3/2\big)$ transitions denoted $\mathrm{\Omega^{\pm 2}_{E2}}$.  The electric quadrupole matrix element is well known from theory and is given in Table \ref{tab:relativesizes}.  The transitions are driven by both the $\mathrm{\hat{x}'}$ and $\mathrm{\hat{y}'}$ components of the laser's electric field but coupling through M1 is suppressed by virtue of its angular momentum selection rule,
 
\begin{equation}
\mathrm{\Omega^{\pm 2}_{E2} = }\frac{k}{4\sqrt{6}\hbar }\frac{1}{\sqrt{5}}\mathrm{E2\,\big|\sin(2\theta)\,E_{x^{'}} \mp 2}\,i\,\mathrm{\sin(\theta)\,E_{y^{'}}\big|  }
\end{equation}

\noindent When driven with vertically polarized light (electric field parallel to Y$^{'}$-axis) and $\theta \sim 90^{\circ}$ both $\mathrm{\Omega^{\pm 2}_{E2}}$ are maximized and flat to leading order in $\theta$.  A precise determination the 2051 nm field amplitude from $\mathrm{\Omega^{\pm 2}_{E2}}$ would thus be relatively straightforward granted that the whole 2051 nm field amplitude can accurately be aligned in the vertical polarization state.  A noteworthy alternative is to measure both $\mathrm{\Omega^{+ 2}_{E2}}$ and $\mathrm{\Omega^{- 2}_{E2}}$ with the 2051 nm beam circularly polarized.  With $\theta$ known these can yield the individual components of the 2051 nm field, $E_{x^{'}}$ and $E_{y^{'}}$.  Measuring the components separately could be a useful check that pure circular polarization was achieved at the ion. 

\begin{figure}[b]
\includegraphics[scale=0.14]{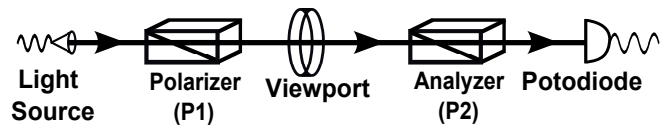} \\
\caption{A schematic of the apparatus used to characterize the stress-optical behavior of the test viewport. }
\label{fig:testchamber}
\end{figure}

\section{Effect of stress induced birefringence on laser polarization}
 \indent Precise and accurate control of the 2051 nm laser polarization will be of general concern to all aspects of the proposed measurement and is paramount to any trapped ion PNC measurement.  While the polarization state of the 2051 nm beam can be controlled outside of the trap, stress induced birefringence in the viewports will tend to unpredictably distort the beam's polarization.  In effect, each point on the viewport acts like a wave-plate with relative phase retardance $\mathrm{\Gamma}$ and an unknown optical axis orientation $\mathrm{\alpha}$, which can be either of the fast or slow axes.  Generally the relative phase difference seen by a beam of wavelength $\mathrm{\lambda}$, that transmits through a transparent isotropic material under mechanical stress, is described by the stress-optic law \cite{Dally78}: 
 
\begin{equation}\label{eq:SIB}
\mathrm{\Gamma} =\frac{2\pi C t}{\lambda}\big(\sigma_{11}-\sigma_{12}\big)
\end{equation} 

\noindent where C is the stress-optics coefficient, t is the thickness of the sample, and $\sigma_{11}$ and $\sigma_{22}$ are first and second principle stresses, respectively.  Distortions of this type need to be limited to a few percent to make possible accurate determination of $\mathrm{\Omega_{M1}}$ and the 2051 nm laser field strength.  A direct measurement of the effect is difficult because the trapping apparatus is embedded within a vacuum system where the beam cannot be easily accessed.  In order to estimate the size of the effect we have measured the amount of stress induced birefringnece in a single test viewport.

\begin{figure}[t]
\includegraphics[scale=0.26]{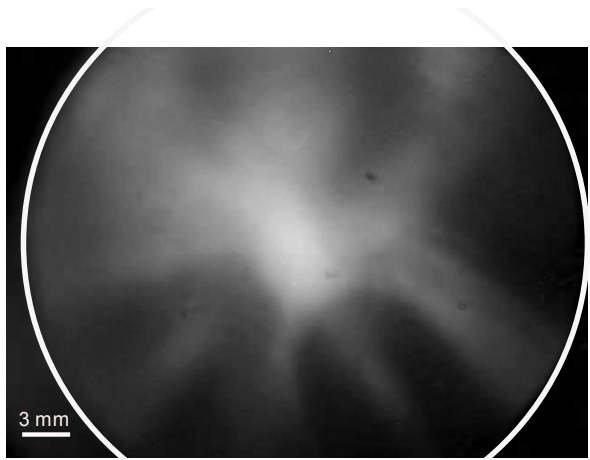}
\caption{A CCD image of the stress pattern in the viewport obtained by shining incoherent red light through the viewport between a pair of crossed polarizers.  The image shows the stress-optic behavior of the viewport qualitatively.  Bright regions in the image correspond to higher amounts of stress induced birefringence.  The viewport edge is indicated in white.}
\label{fig:ios}
\end{figure}

\indent A standard 1.33 inch diameter fused silica viewport from MDC Vacuum Products, LLC served as a test viewport.  To replicate a viewport \emph{in situ}, the test viewport was baked for five days at a temperature of 150$^{\circ}$ C while pumped down to a pressure of $\mathrm{\sim 2.0 \times 10^{-7}}$ Torr. Although longer bakes and lower pressures are needed for real trapped ion experiments these were sufficient for our intentions here.  A charge-coupled device (CCD) image of the stress induced birefringence present in the viewport was taken prior to baking and is shown in Fig. \ref{fig:ios}.  Similar images were taken of previously used viewports and suggest that comparable amounts of stress were present and so we expect the results reported here to be typical in magnitude.

\begin{figure}[ht]
\includegraphics[scale=0.33]{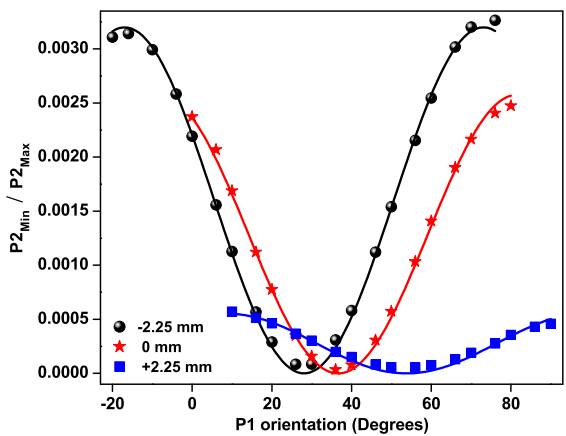}
\caption{(Color online) Ratio of $\mathrm{P_{Min}}$ to $\mathrm{P_{Max}}$ as a funtion of P1 orientation at the center and $\pm$2.25 mm off center of the viewports.  The data is fit to Eq. (\ref{eq:ratio}) to get $\mathrm{\Gamma}$ and $\mathrm{\alpha}$ for each spatial incidence.}
\label{fig:ellipticity3}
\end{figure}

\indent To quantify the effect we have determined the orientation of the viewport's optical axes, $\mathrm{\alpha}$, and the relative phase retardation between the axes, $\mathrm{\Gamma}$, at its center and two points on a diameter.  The measurements were made with a linearly polarized 650 nm beam with a full width at half max of 2 mm.  The choice to take the measurements at this wavelength was made simply for our convenience.  A schematic of the experiment is provided in Fig. \ref{fig:testchamber}.  The 650 nm light was delivered via single mode optical fiber to a Glan-Thompson polarizer (P1) that controlled the linear polarization angle $\mathrm{\Theta}$ of the light incident on the front of the viewport.  We take $\mathrm{\Theta}=0$ to mean horizontal polarization.  A second identical Glan-Thomson polarizer (P2) is placed after the test viewport and was used to analyze the beam's polarization state after transmitting through the viewport.  The Glan-Thompson polarizers had a manufacturer quoted extinction ratio of about 100 000:1 in power which was verified.  The relative optical power after P2 was measured with a silicon photodiode.  The optical fiber, polarizers, and photodiode were mounted on a translation stage that was moved transverse to the viewport with a micrometer.  \\
\indent Each measurement consisted of placing P1 to a known orientation and then rotating P2 to find the minimum and maximum power in the beam, $\mathrm{P_{Min}}$ and $\mathrm{P_{Max}}$.  The ratio $\mathrm{P_{Min}} / \mathrm{P_{Max}}$ is approximately the square of the ellipticity introduced into the beam's polarization by the viewport.  The finite extinction ratio of P1 and P2 did cause a non-zero $\mathrm{P_{Min}} / \mathrm{P_{Max}}$ without a viewport placed between the polarizer, however this contribution was ten times smaller than what was caused by the viewport and so was ignored.

\indent The optical axis orientation and phase retardation of the viewport at a given location are related to $\mathrm{P_{Min}} / \mathrm{P_{Max}}$ by,

\begin{equation}\label{eq:ratio}
\frac{\mathrm{P_{Min}}}{\mathrm{P_{Max}}} =\frac{\mathrm{\sin}^2\big[\frac{\mathrm{\Gamma}}{2}\big] \,\mathrm{\sin}^2\big[2(\mathrm{\Theta}-\mathrm{\alpha})\big]}{1-\mathrm{\sin}^2\big[\frac{\mathrm{\Gamma}}{2}\big]\, \mathrm{\sin}^2\big[2(\mathrm{\Theta}-\mathrm{\alpha})\big]}
\end{equation} 

\noindent To find $\mathrm{\Gamma}$ and $\alpha$ we measured the power ratio for 18 orientations of P1 between $0\,^{\circ}$ to $90\,^{\circ}$ at each spatial incidence, as shown in Fig \ref{fig:ellipticity3}.  The parameters $\mathrm{\Gamma}$ and $\mathrm{\alpha}$ were found by fitting Eq. \ref{eq:ratio} to the data, the results of which are displayed in Table \ref{tab:fit}.

\begin{table}[ht]
\centering
\begin{tabular}{|c|c|c|}
\hline
Position&\multicolumn{2}{c|}{Measured at 650 nm}\\ 
&$\mathrm{\Gamma}$&$\mathrm{\alpha}$ \\
\hline
 \;\,-2.25 mm& $6.47\,^{\circ} \pm 0.02\,^{\circ}$ & $28.08\,^{\circ} \pm 0.15\,^{\circ}$ \\ [1ex]
 \;\;\,0.00 mm& $5.81\,^{\circ} \pm 0.02\,^{\circ}$ & $39.67\,^{\circ} \pm 0.18\,^{\circ}$ \\ [1ex]
+2.25 mm & $2.69\,^{\circ} \pm 0.05\,^{\circ}$ & $53.79\,^{\circ} \pm 0.57\,^{\circ}$ \\ [1ex]
\hline
\end{tabular}
\caption {The retardance and optical axis orientation of the viewport for the center and opposite points along the edge.}
 \label{tab:fit} 
\end{table}

\indent  These measurements place upper bounds on the anticipated polarization distortion of the 2051 nm beam.  The stress-optic law predicts that $\mathrm{\Gamma}$ will be about three times less for a 2051 nm beam.  The effect is further suppressed at 2051 nm by dispersion of the stress-optic coefficient \cite{Vasudevan72}, although it is unknown by how much.  At the center of the viewport, not accounting for dispersion, a wave-plate retardance of $\sim 2^{\circ}$ can be expected for a 2051 nm beam.  If left uncorrected this value would shift M1 by about 5\%.  This indicates that distortions in 2051 nm polarization from stress induced birefringence can likely be tolerated for the M1 measurement.  However these distortions will need to be further suppressed or compensated for in a trapped ion PNC measurement.    

\section{Conclusion}
A parity nonconservation experiment with single trapped Ba$^{+}$ requires the magnetic dipole transition moment for the 2051 nm $6S_{1/2} \leftrightarrow 5D_{3/2}$ transitions be known.  To date, there is one calculation of M1 and it predicts a value dominated by electron-electron correlation effects \cite{Sahoo06}.  We have therefore proposed an approach for extracting M1 from a measurement of $\mathrm{\Omega_{M1}}$ using an E2-M1 intensity interference.  In our approach the relative phase between $\mathrm{\Omega_{E2}}$ and $\mathrm{\Omega_{M1}}$ is controlled experimentally for the $6S_{1/2}\big( \mathrm{m}=-1/2\big) \leftrightarrow 5D_{3/2}\big( \mathrm{m}=-1/2\big)$ transition so that $\mathrm{\Omega_{E2}}$ can be eliminated to reveal $\mathrm{\Omega_{M1}}$. 

\indent We describe two versions of the measurement using the $\mathrm{\Delta m=0}$ transition.  In one version the 2051 nm beam alignment is controlled by rotating the ion's quantization axis symmetrically about 90$^{\circ}$.  A second approach is to drive the transition with either sense of circular polarization, which has the equivalent effect.  These two approaches suffer from different systematics and can therefore be used to check for consistency.  To extract M1 from $\mathrm{\Omega_{M1}}$ the 2051 nm beam alignment and intensity need be known.  We suggest that these can be had from measurements of the Rabi frequencies for transitions to particular $5D_{3/2}$ states with judiciously selected 2051 nm beam polarizations.  The feasibility of the measurement, in either approach, therefore depends critically on the ability to carefully control the 2051 nm beam's polarization.

\indent A general concern to the proposed measurement is the effect that stress induced birefringece will have on the polarization of a 2051 nm beam.  To estimate how much polarization distortion can be expected we have measured the effect in a test viewport.  Unfortunately it is difficult to measure the effect directly with the 2051 nm beam so we have used 650 nm beam instead.  Stress induced biregringence falls off faster than 1/$\mathrm{\lambda}$ and accordingly our measurements place an upper bound for what will be found with the 2051 nm beam.  The result of these measurements suggest that the polarization distortion to be expected in a 2051 nm beam will be small enough to be insignificant to the M1 measurement. \\

\section{Acknowledgments}
\indent The authors would like to thank the other members of the Blinov group; Richard Graham, Zichao Zhou, John Wright, Tomasz Sakrejda, Thomas Noel, Carolyn Auchter, Chen-Kuan Chou and Alexander Sivitilli for discussion and commentary.  The authors would also like to thank Alan Jamison and Benjamin Plotkin-Swing from the University of Washington's Ultra-Cold Atoms group for a helpful discussion at an early stage of this work.  This work was supported by the National Science Foundation, Grant No. PHY-09-06494.
    

\end{document}